%% file: main.tex
\documentclass[aps,prd,citeautoscript,twocolumn,groupedaddress,amsfonts,amssymb,amsmath,showpacs,showkeys,nofootinbib,nolongbibliography]{revtex4-2}

\usepackage{graphicx}


\begin{document}


\title{A comparison between the deflection angles 
of massive and massless particles in the 
Shchwarzschild space-time and their consequences on black hole shadows}

\author{S. Mendoza}
\email[Email address: ]{sergio@astro.unam.mx}
\author{M. Santiba\~nez}
\email[Email address: ]{msantibanez@astro.unam.mx}
\affiliation{Instituto de Astronom\'{\i}a, Universidad Nacional
                 Aut\'onoma de M\'exico, AP 70-264, Ciudad de M\'exico, 04510,
	         M\'exico \\
            }

\date{\today}

\begin{abstract}
  We present comparisons of the deflection angles 
of massless and massive particles in the Schwarzschild space-time.  For the case of photons in a general static space-time,
we construct a spatial 3D equation of motion for their path that leads to an implicit formula for the deflection angle.  We then compare our results
with well known results of the literature in the 
Schwarzschild space-time.
For the case of massive particles we calculate the
deflection angle only in the Schwarzschild space-time.  The end result is that as the velocity of 
any massive particle diminishes, the deflection angle increases.  To show the relevance of these 
comparisons, we constructed different black hole
shadows for massive particles in order to be compared with a shadow made by of photons.
\end{abstract}


\keywords{General relativity and gravitation; 
Black holes}

\maketitle

\section{Introduction}
\label{Introduccion}

Since \citet{eddington1919} obtained evidence for the deflection of light generated by our Sun, telescopes had caught pictures that show how light beams are affected by gravitational fields generated by mass distributions, as we can see in some images of Hubble and James Webb telescope.

\begin{figure}
    \centering
    \includegraphics[scale=0.4]{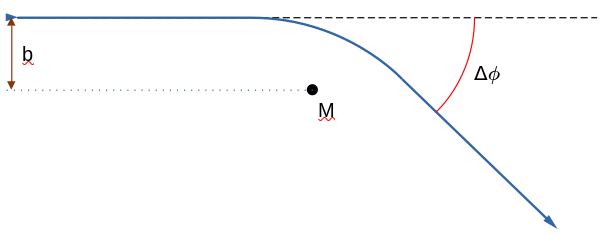}
    \caption{The diagram shows the trajectory of a light beam in blue.  This path starts from an infinite distance to the left with respect to the mass \( M \), at  an impact parameter $b$. The trajectory is deflected an angle $\Delta\phi$ by the gravitational field produced by the mass $M$.}
    \label{fig:first_DA_diagram}
\end{figure}

The first expression which described the deflection angle of light beams passing close to a massive point source \( M \) was found by Soldner (1802) (see e.g. \citet{Soldner1802} for
an excellent review on this).  This Newtonian formula for the deflection angle of a test particle is given by:

\begin{equation}
    \Delta\phi_{\text{Soldner}}=\dfrac{2GM}{v^2b},
\label{eq:Soldner}
\end{equation}

\noindent where $v$ is the velocity of the test particle, $G$ is Newton's gravitational constant and $b$ is the impact parameter shown in Figure~\ref{fig:first_DA_diagram}. Since the mass of the test particle does not appear on equation~\eqref{eq:Soldner}, the original idea of this Soldner deflection angle, was that it would predict also the deflection of massless photon test particles  passing close to the gravitational field of the massive object \( M \), by simply assuming that the velocity \( v \) is the velocity of light \( c \).

 Using general relativistic arguments, \citet{einstein1916} calculated the correct
expression for the small deflection angle of light beams passing close to a weak 
gravitational field produced by a point mass source \( M\):

\begin{equation}\label{eq:Einstein}
    \Delta\phi_{\text{Einstein}}=\dfrac{4GM}{c^2b},
\end{equation}

\noindent which is twice the deflection angle calculated by Soldner.  

 Over the years, the interaction of light rays with gravitational fields have become 
a very active research subject with plenty of astrophysical applications.  Among these 
are the first shadows generated by a black hole \citep{Luminet1978} and the observations by the Event Horizon Telescope (EHT) \citep{Event_Horizon_Telescope_Collaboration_2022}, and the extensive studies of gravitational 
lenses~\citep[e.g.][]{Schneider2012gravitational,dodelson2017gravitational}.

The study of the deflection of massive relativistic particles passing close to a weak gravitational field has been studied using different variational techniques~\citep[][]{A_Accioly_2002,Pang_2019,Li_2019}. Neutrinos are relativistic massive particles~\citep{A_Modern_Introduction_to_Neutrino_Physics}, difficult to detect~\citep{Icecube} but they follow geodesics on a curved space-time.  In fact, since neutrinos could be emitted by an accretion disc of 
a compact object~\citep{Rebecca_Surman_et.al.,PhysRevD.47.5270}, one could in principle draw the shadow of 
neutrino particles emitted from a black hole.

  The article is organised as follows.  In Section~\ref{Sec:Motion equations of a light beam}, we obtain a new set of 3D equations for the path of a light beam valid for any stationary space-time and we test the results in Section~\ref{Sec:Deflection angle of a light beam in Schwarzschild space-time}
by calculating the deflection angle for a Schwarzschild space-time.
In Section~\ref{Sec:Deflected angle for massive particles} we performed a similar analysis for ultrarelativistic massive particles, finding deflection angles on different ways. Finally, we evolved the light motion equations obtained and the geodesic equation of a massive particle using the free GNU Public Licensed software aztekas-shadows \copyright 2024 Magallanes, Mendoza \& Santiba\~nez that will be publicly available in the near future, to generate Schwarzschild black hole shadows of massless and massive particles.

\section{3D motion equations of a light beam}
\label{Sec:Motion equations of a light beam}

  On any four-dimensional pseudo-Riemannian manifold adapted to general relativity, the 
four-interval \( \mathrm{d} s \) as a function of the metric \( g_{\mu\nu} \) can be decomposed
as follows \citep[see e.g.][]{landau2013classical}:

\begin{equation}\label{eq:4D_metric}
\begin{aligned}
\mathrm{d}s^2&=g_{\mu\nu}\mathrm{d}x^{\mu}\mathrm{d}x^{\nu}, \\
&=g_{00}(\mathrm{d}x^0)^2+2g_{0i}\mathrm{d}x^0\mathrm{d}x^i+g_{ik}\mathrm{d}x^i\mathrm{d}x^k, \\
&=\left(\sqrt{g_{00}}~\mathrm{d}x^0+\dfrac{g_{0i}}{\sqrt{g_{00}}}~\mathrm{d}x^i\right)^2-\mathrm{d}l^2,
\end{aligned}
\end{equation}

\noindent where \( x^\mu = ( x^0,\ x^1,\ x^2,\ x^3 )\) represent four coordinates, $\mathrm{d}l^2 
:= \gamma_{ik}\mathrm{d}x^i\mathrm{d}x^k$ is the spatial 3D-arch length and:

\begin{equation}
\gamma_{ik}\mathrel{\mathop:}=\left(-g_{ik}+\dfrac{g_{0i}g_{0k}}{g_{00}}\right),
\end{equation}

\noindent represents the spatial 3D-metric. In the previous equations and in what follows Greek indices take space-time values 
(\( 0,1,2,3 \)) and Latin ones take spatial values (\(1,2,3\)).  We use Einstein's summation convention
over repeated indices and we use a (\(+,-,-,-\)) signature for the metric \citep[see e.g.][]{misner2017gravitation}.

  Light trajectories occur in null \( \mathrm{d} s = 0 \) geodesics and so, using 
  equation~\eqref{eq:4D_metric}, it turns out that:

\begin{equation}\label{eq:time_interval}
\begin{aligned}
\mathrm{d}x^0&=-\dfrac{g_{0i}}{g_{00}}\mathrm{d}x^i\pm\dfrac{1}{\sqrt{g_{00}}}\mathrm{d}l, \\
&=\dfrac{1}{g_{00}}\left[-g_{0i}\dfrac{\mathrm{d}x^i}{\mathrm{d}l}\pm\sqrt{g_{00}}\right]\mathrm{d}l.
\end{aligned}
\end{equation}

\noindent The common definition of the refraction index is given by~\citep{Poisson2014gravity}:

\begin{equation}\label{eq:refraction_index}
n\mathrel{\mathop:}=\dfrac{c}{\mathrm{d}l/\mathrm{d}t}=\dfrac{1}{g_{00}}\left[-g_{0i}e^i+\sqrt{g_{00}}\right],
\end{equation}

\noindent with the positive sign for a time evolution to the future and so,
we can rewrite equation~\eqref{eq:time_interval} as:

\begin{equation}\label{eq:fermat_princ_1}
\mathrm{d}x^0=n\mathrm{d}l,
\end{equation}

\noindent where $e^i := \mathrm{d}x^i/\mathrm{d}l$ is a tangent vector to the space 3D-path of the 
light beam.  Note that since

\begin{equation}
e_i=\gamma_{ik}e^k, \qquad \text{and so} \qquad e^i=\gamma^{ik}e_k,
\label{econtracov}
\end{equation}

\noindent then:

\begin{equation}
e_ie^i=\gamma_{ik}e^ke^i=\gamma_{ik}\dfrac{\mathrm{d}x^k}{\mathrm{d}l}\dfrac{\mathrm{d}x^i}{\mathrm{d}l}=\dfrac{\gamma_{ik}\mathrm{d}x^k\mathrm{d}x^i}{\mathrm{d}l^2}=1,
\end{equation}

\noindent i.e., $e^i$ is a unit vector.

  For the development of this article it will be very important Fermat's principle, 
which can be expressed simply as \emph{`The minimum possible time it takes for light to travel a path between two points in space'}, and can be interpreted as an extremum relativistic principle in time as follows~\citep{Schneider2012gravitational}:

\textit{Consider an event $A$ where the radiation is emitted and a time-like world-line $\mathit{l}$ 
for which $\mathrm{d}s^2>0$. A ray of light is a soft null curve \( \Gamma \) that propagates from $A$ through
$\mathit{l}$ if and only if the arrival time $t$ on $\mathit{l}$ is stationary under 
first order variations of  \( \Gamma \) within the set of soft null curves from $A$ through $\mathit{l}$, in other words
}

\begin{equation}
  \delta \int{ \mathrm{d} t } = \delta t=0,
\end{equation}

\noindent which, according to equation~\eqref{eq:fermat_princ_1} yields:

\begin{equation}\label{eq:fermat_princ_2}
\delta\int n~\mathrm{d}l=0.
\end{equation}

  In here and in what follows we will only deal with \emph{stationary} space-times in the sense 
defined by \citet{landau2013classical}\footnote{As described by \citet{landau2013classical}, a 
constant gravitational field is one for which there exists a system of reference where the metric tensor components do not depend on the time coordinate. Furthermore, a constant gravitational
field for which $g_{0i}\neq 0$ for some $i$ is called stationary.  If  $g_{0i}=0$, for all \( i \) then the gravitational field is called static.}.  For such space-times
equation~\eqref{eq:fermat_princ_2} with a general refraction index takes the form:

\begin{equation}\label{eq:fermat_princ_3}
\delta x^0=\int\left[\dfrac{\partial n}{\partial x^k}~\delta x^k~\mathrm{d}l+n~\delta\left(\mathrm{d}l\right)\right]=0.
\end{equation}	

\noindent Since the null variations of $\mathrm{d}l$ are given by:
\begin{equation}
\begin{aligned}
\delta\left(\mathrm{d}l\right)&=\delta\sqrt{\gamma_{ik}\mathrm{d}x^i\mathrm{d}x^k}, \\
&=\left[\dfrac{1}{2}\dfrac{\partial\gamma_{ij}}{\partial x^k}e^ie^j\right]\delta x^k\mathrm{d}l+e_k\mathrm{d}\left(\delta x^k\right),
\end{aligned}
\end{equation}

\noindent equation~\eqref{eq:fermat_princ_3} can be integrated by parts to yield:
\begin{equation}
\begin{aligned}
\delta x^0=\int &\dfrac{\partial n}{\partial x^k}\delta x^k\mathrm{d}l +\left[ \dfrac{n}{2}\dfrac{\partial \gamma _{ij}}{\partial x^k}e^ie^j\right] \delta x^k\mathrm{d}l \\
&-\left[ \dfrac{\partial n}{\partial x^i}e^ie_k+n\dfrac{\mathrm{d}e_k}{\mathrm{d}l }\right] \mathrm{d}l \delta x^k, \\
=\int &\left[ \dfrac{\partial n}{\partial x^k}+\dfrac{n}{2}\dfrac{\partial \gamma _{ij}}{\partial x^k}e^ie^j\right. \\
&\left.-\dfrac{\partial n}{\partial x^i}e^ie_k-n\dfrac{\mathrm{d}e_k}{\mathrm{d}l }\right] \mathrm{d}l~\delta x^k=0.
\end{aligned}
\end{equation}

\noindent Since the previous integral is null for any variation \( \delta x^k \), it then follows that:  

\begin{equation}\label{eq:de_k/dl}
\dfrac{\mathrm{d}e_k}{\mathrm{d}l }=\dfrac{\partial \ln n }{\partial x^k}+\dfrac{1}{2}\dfrac{\partial \gamma _{ij}}{\partial x^k}e^ie^j-\dfrac{\partial 
\ln n }{\partial x^i}e^ie_k,
\end{equation}

\noindent or by using equation~\eqref{econtracov}:

\begin{equation}\label{eq:de^k/dl}
\begin{aligned}
\dfrac{\mathrm{d}e^q}{\mathrm{d}l }=&\left[ \gamma ^{kq}\dfrac{\partial \ln n}{\partial x^k}-\dfrac{\partial \ln n}{\partial x^i}e^ie^q\right] \\
&+\gamma ^{kq}\left[ \frac{1}{2}\dfrac{\partial \gamma _{ij}}{\partial x^k}e^ie^j-\dfrac{\partial \gamma _{kp}}{\partial x^m}e^me^p\right].
\end{aligned}
\end{equation}

\noindent Direct substitution of equation~\eqref{eq:refraction_index} into the previous equation yields:

\begin{equation}\label{eq:general_evolution_equations}
\begin{aligned}
\dfrac{\mathrm{d}e^q}{\mathrm{d}l }=\dfrac{\gamma^{kq}}{n}&\left\lbrace \dfrac{1}{g_{00}}\left[\dfrac{1}{2\sqrt{g_{00}}}\dfrac{\partial g_{00}}{\partial x^k}-e^i\dfrac{\partial g_{0i}}{\partial x^k}\right]\right. \\
&\left.-\dfrac{1}{g_{00}^2}\dfrac{\partial g_{00}}{\partial x^k}\left(\sqrt{g_{00}}-g_{0i}e^i\right)\right\rbrace \\
-\dfrac{e^ie^q}{n}&\left\lbrace \dfrac{1}{g_{00}}\left[\dfrac{1}{2\sqrt{g_{00}}}\dfrac{\partial g_{00}}{\partial x^i}-e^m\dfrac{\partial g_{0m}}{\partial x^i}\right]\right. \\
&\left.-\dfrac{1}{g_{00}^2}\dfrac{\partial g_{00}}{\partial x^i}\left(\sqrt{g_{00}}-g_{0m}e^m\right)\right\rbrace \\
+\gamma ^{kq}&\left[ \frac{1}{2}\dfrac{\partial \gamma _{ij}}{\partial x^k}e^ie^j-\dfrac{\partial \gamma _{kp}}{\partial x^m}e^me^p\right].
\end{aligned}
\end{equation}

\noindent This expression represents the rate of change of the vector $e^i$ between two points
along the path of a null geodesic, as a function only of the components of the 3-metric, its 
derivatives and the vector $e^i$.  In other words, it represents the 3D equation of motion of a light beam.

\section{Deflection angle of a light beam in Schwarzschild space-time}
\label{Sec:Deflection angle of a light beam in Schwarzschild space-time}

Let us consider a spherically symmetric coordinate system.  In Schwarzschild-like coordinates  (\(ct,r,\theta,\varphi\)), the interval is given by~\citep{hobson2006general}:

\begin{equation}\label{metrica_sim_esf}
\mathrm{d}s^2=g_{00}c^2\mathrm{d}t^2+g_{rr}\mathrm{d}r^2-r^2\mathrm{d}\Omega,
\end{equation}

\noindent where \( \mathrm{d}\Omega \) is the angular displacement and the components $g_{00}$ and $g_{rr}$ are functions that depend only on the \( r \) coordinate.

In this coordinate system, equations~\eqref{eq:general_evolution_equations} are written as:

\begin{equation}\label{eq:evol_spheric_system}
\begin{aligned}
\dfrac{\mathrm{d}^2r}{\mathrm{d}l^2}=&-\dfrac{1}{g_{rr}}\left[\dfrac{\mathrm{d}\ln n}{\mathrm{d}r}+\dfrac{1+g_{rr}\left(e^r\right)^2}{r}+\dfrac{\left(e^r\right)^2}{2}\dfrac{\mathrm{d}g_{rr}}{\mathrm{d}r}\right] \\
&-\left(e^r\right)^2\dfrac{\mathrm{d}\ln n}{\mathrm{d}r}, \\
\dfrac{\mathrm{d}^2\theta}{\mathrm{d}l^2}=&\left(e^{\varphi}\right)^2\sin\theta\cos\theta-\dfrac{2e^re^{\theta}}{r}-e^re^{\theta}\dfrac{\mathrm{d}\ln n}{\mathrm{d}r}, \\
\dfrac{\mathrm{d}^2\varphi}{\mathrm{d}l^2}=&-2e^{\theta}e^{\varphi}\cot\theta-\dfrac{2e^re^{\varphi}}{r}-e^re^{\varphi}\dfrac{\mathrm{d}\ln n}{\mathrm{d}r}.
\end{aligned}
\end{equation}

\noindent with $n=1/\sqrt{g_{00}}$, i.e., $n=n(r)$. These are a system of coupled second order differential 
equations.

\begin{figure}
    \centering
    \includegraphics[scale=0.33]{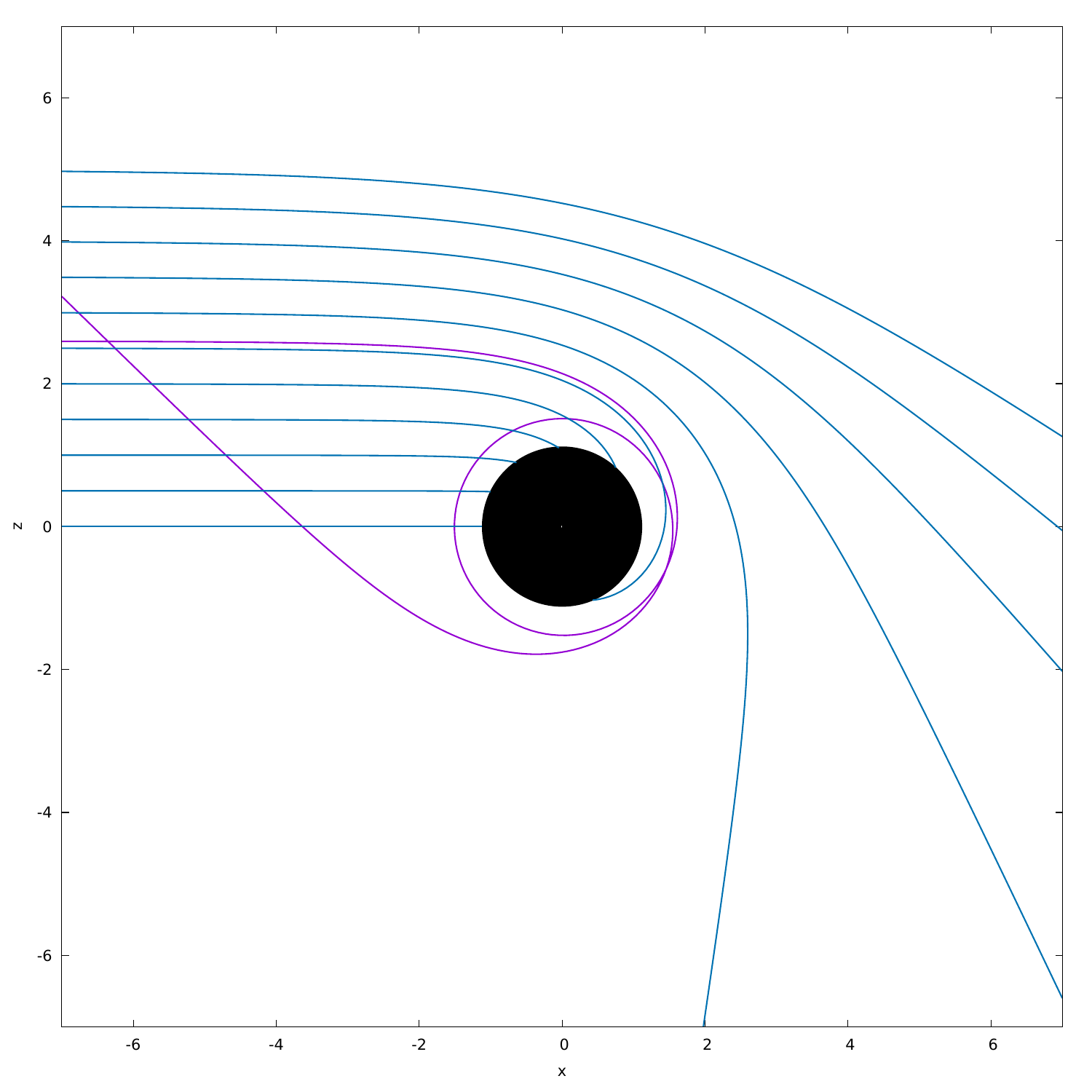}
    \caption{The Figure shows different light beam trajectories in blue deflected by a Schwarzschild black hole (represented as a black circle). All trajectories start horizontally far to the left of the figure at position \( x_0 = -10^4 r_\mathrm{S} \) with an initial tangent vector $e^x=1$, $e^y=e^z=0$ and a given impact parameter $b$.  Each trajectory 
    is the numerical 4th order Ruge-Kutta solution to the 3D-equations of motion of a null geodesic represented in equation~\eqref{eq:evol_spheric_system}.
    The coordinates in the Figure are normalized to the Schwarzschild radius $r_\mathrm{S}$.}
    \label{fig:light_beam_paths}
\end{figure}

  For the Schwarzschild space-time the metric tensor takes the following form~\citep[e.g.][]{hobson2006general}:

\begin{equation}
g_{00}=1-\frac{r_\mathrm{S}}{r}\hspace{1cm}g_{rr}=\left(1-\frac{r_\mathrm{S}}{r}\right)^{-1},
\end{equation}

\noindent where $r_\mathrm{S}=2GM/c^2$ is the Schwarzschild radius. With this change of variable and knowing the initial values of the position $x^i$ and the tangent vector $e^i$ for a given massless particle, we used a 4th order Runge-Kutta method to
integrate numerically  equations~\eqref{eq:evol_spheric_system} for any $r>r_\mathrm{S}$ with a given
impact parameter $b$. The results are presented in   Figure~\ref{fig:light_beam_paths}, with light beams 
emitted so far to the left of the diagram ($r_\text{initial} = 10^4 r_\mathrm{S}$), where the space-time is essentially flat~\citep[][]{Ziri_Younsi}.

\begin{figure}
    \centering
    \includegraphics[scale=0.30]{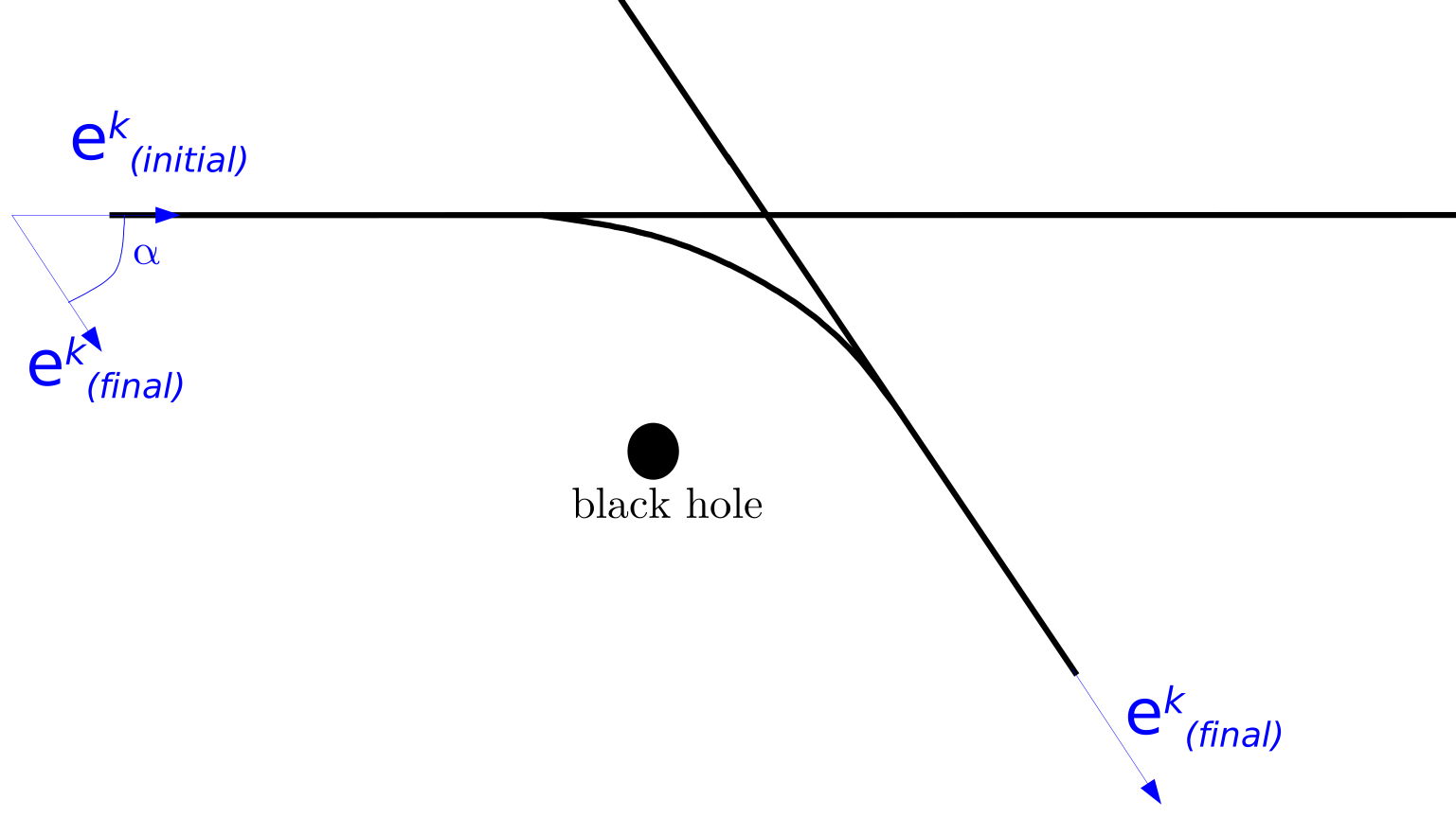}
    \caption{The diagram shows the vectors $e^k_\mathrm{(initial)}$ and $e^k_\mathrm{(final)}$ . The angle $\alpha$ is defined as the angle betwen two vectors in flat space-time.}
    \label{fig:deflection_with_ei_ef}
\end{figure}

If we analyze impact parameters greater than $4r_\mathrm{S}$ we will avoid trajectories that complete one or more turns around the black hole (and the light beams which cross the event horizon). For this kind of trajectories we can define two 
tangent vectors to the trajectory as we can see in Figure~\ref{fig:deflection_with_ei_ef}: an initial one given by $e^k_\text{(initial)}$ and a final one  $e^k_\text{(final)}$. These two vectors are too far that space-time is essentially flat around them. Under these considerations the deflection angle can thus be calculated using the standard trigonometric cosine law as~\citep[][]{2014PhDT_Younsi}:

\begin{equation}\label{eq:alpha_path_integral}
\begin{aligned}
    \alpha=\cos^{-1}\dfrac{e^k_\text{(initial)} e_{ k\text{(final) }} } {| e^k_\text{(initial)} | 
    \, | e^k_\text{(final) } | 
    }.
\end{aligned}
\end{equation}

\noindent In the weak field limit of approximation, this deflection angle converges to the Einstein deflection angle~\eqref{eq:Einstein}.  

  Equation~\eqref{eq:alpha_path_integral} represents an alternative expression to the deflection angle  \( \beta \) 
presented by \citet{Keeton2005Petters}:

\begin{equation}\label{eq:beta_keeton}
\beta = 2\int _{r_0} ^{\infty }\dfrac{[-g_{00}(r)g_{11}(r)]^{1/2}}{r\left[\left(\dfrac{r}{r_0}\right)^2 g_{00}(r_0)-g_{00}(r)\right]^{1/2}}\mathrm{d}r-\pi.
\end{equation}

\begin{figure}
    \centering
    \includegraphics[scale=0.34]{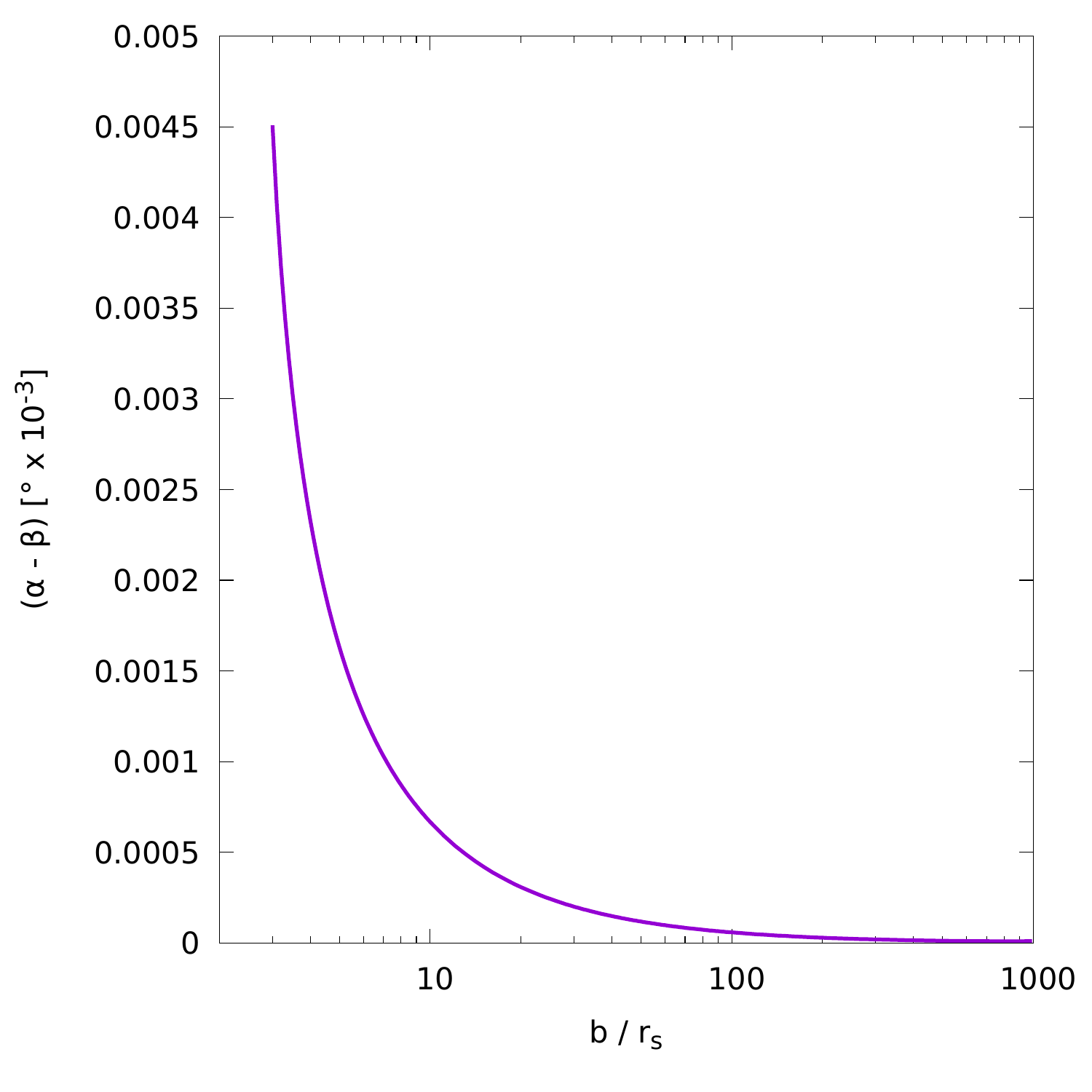}
    \caption{Difference between the deflection angle \( \alpha \) calculated with the
    formula~\eqref{eq:alpha_path_integral} and the numerical value for \( \beta \) in equation~\eqref{eq:beta_keeton} as a function of the impact parameter \( b \) measured in units of the gravitational radius \( r_\text{S} \).}
     
    \label{fig:dif_path_keeton}
\end{figure}

 A numerical comparison between $\alpha$ and $\beta$ as function of the impact parameter is shown in 
Figure~\ref{fig:dif_path_keeton}.  The integral~\eqref{eq:beta_keeton} was computed with the 
help of the GNU Scientific Library (GSL)~\citep{GNU_Scientific_Libraries} since that integral
presents a pole at \( r = r_0 \).   The result is that both approaches produce the same
deflection angle.

\section{Deflection angle for massive particles}
\label{Sec:Deflected angle for massive particles}

We can write the equations for the three constants of motion in  Schwarzschild spacetime 
as~\citep[cf.][]{hobson2006general}: 

\begin{equation}\label{eq:cst_Schwarzschild}
\begin{aligned}
k = &\left(1-\dfrac{r_\mathrm{S}}{r}\right)\dfrac{\mathrm{d}t}{\mathrm{d}\tau}, \\
c^2 = &c^2\left(1-\dfrac{r_\mathrm{S}}{r}\right)\left(\dfrac{\mathrm{d}t}{\mathrm{d}\tau}\right)^2 \\
 &-\left(1-\dfrac{r_\mathrm{S}}{r}\right)^{-1}\left(\dfrac{\mathrm{d}r}{\mathrm{d}\tau}\right)^2-r^2\left(\dfrac{\mathrm{d}\varphi}{\mathrm{d}\tau}\right)^2, \\
h = &r^2\dfrac{\mathrm{d}\varphi}{\mathrm{d}\tau},
\end{aligned}
\end{equation}

\noindent where $k$ is proportional to the energy and $h$ is proportional to the 
angular momentum~\citep[][]{Tejeda_and_Aguayo2019}.

If we define $u=1/r$  the system of equations~\eqref{eq:cst_Schwarzschild} can be 
rewritten as:

\begin{equation}
\begin{aligned}\label{eq:for_u}
\dfrac{\mathrm{d}^2u}{\mathrm{d}\varphi^2}+u=\dfrac{GM}{h^2}+\dfrac{3GM}{c^2}u^2
\end{aligned}
\end{equation}

\begin{figure}
    \centering
    \includegraphics[scale=0.23]{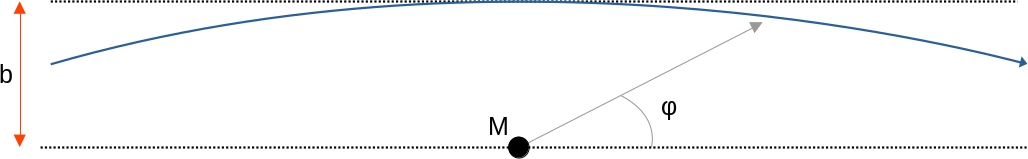}
    \caption{The Figure shows the trajectory of a massive test particle as it moves around a central mass \( M \) with a sufficiently large impact parameter $b$, that produces a small deflection angle on its path. }
    \label{fig:diagram_small_defletion_mass_particle}
\end{figure}

Figure~\ref{fig:diagram_small_defletion_mass_particle} shows  the symmetric trajectory of
a massive test particle as it moves about the coordinate origin where a point mas \( M \) is located.  For sufficiently large impact parameters \( b \), the deflection angle is small with a value $\Delta\varphi=|2\varphi(r=\infty)|$.

The homogeneous solution of equation~\eqref{eq:for_u} is $u_0(\varphi)=\sin\varphi / b$,
and so the first order correction in $\Delta u$ is:

\begin{equation}\label{eq:result_variation_method}
\Delta u=\dfrac{3GM}{2c^2b^2}\left(1+\dfrac{1}{3}\cos(2\varphi)\right)+\dfrac{GM}{h^2}.
\end{equation}

In the limit $r\rightarrow\infty$ and small angle approximation $\sin\varphi\approx\varphi$
which implies $u\rightarrow0$ and $\cos2\varphi\approx 1$ results in an analytic expression
for small deflection angles given by: 

\begin{equation}\label{eq:analytic_deflection_masive}
\Delta\varphi=\dfrac{4GM}{c^2b}+\dfrac{2GM}{c^2b}\left(\dfrac{1-\beta_{\infty}^2}{\beta_{\infty}^2}\right),
\end{equation}

\noindent where $\beta_{\infty}:=v_{\infty}/c$ and $v_{\infty}$ is the velocity of the 
massive particle at infinity where it was emited. This result is coherent with the previous one obtained by \citet[][]{A_Accioly_2002,Pang_2019,Li_2019} to first order in $1/b$.

If we take the ultrarelativistic limit $\beta_{\infty}\rightarrow 1$, equation \eqref{eq:analytic_deflection_masive} becomes Einstein's deflection angle presented in relation~\eqref{eq:Einstein}.  For the non-relativistic limit, where $\beta_{\infty} \ll 1$,
the deflection angle converges to the  Soldner angle shown in~\eqref{eq:Soldner}.

 In general terms, the deflection can be obtained using the system of equations~\eqref{eq:cst_Schwarzschild} by integrating the following expression:

\begin{equation}\label{eq:deflection_masive_part}
\dfrac{\mathrm{d}\varphi}{\mathrm{d}u}=-\dfrac{bV_{\infty}}{\sqrt{V_{\infty}^2+ur_\mathrm{S}+b^2V_{\infty}^2u^2(ur_\mathrm{S}-1)}},
\end{equation}

\noindent where $V_{\infty}=\beta_{\infty}/\sqrt{1-\beta_{\infty}}$.

\begin{figure*}
    \centering
    \includegraphics[width=0.45\textwidth]{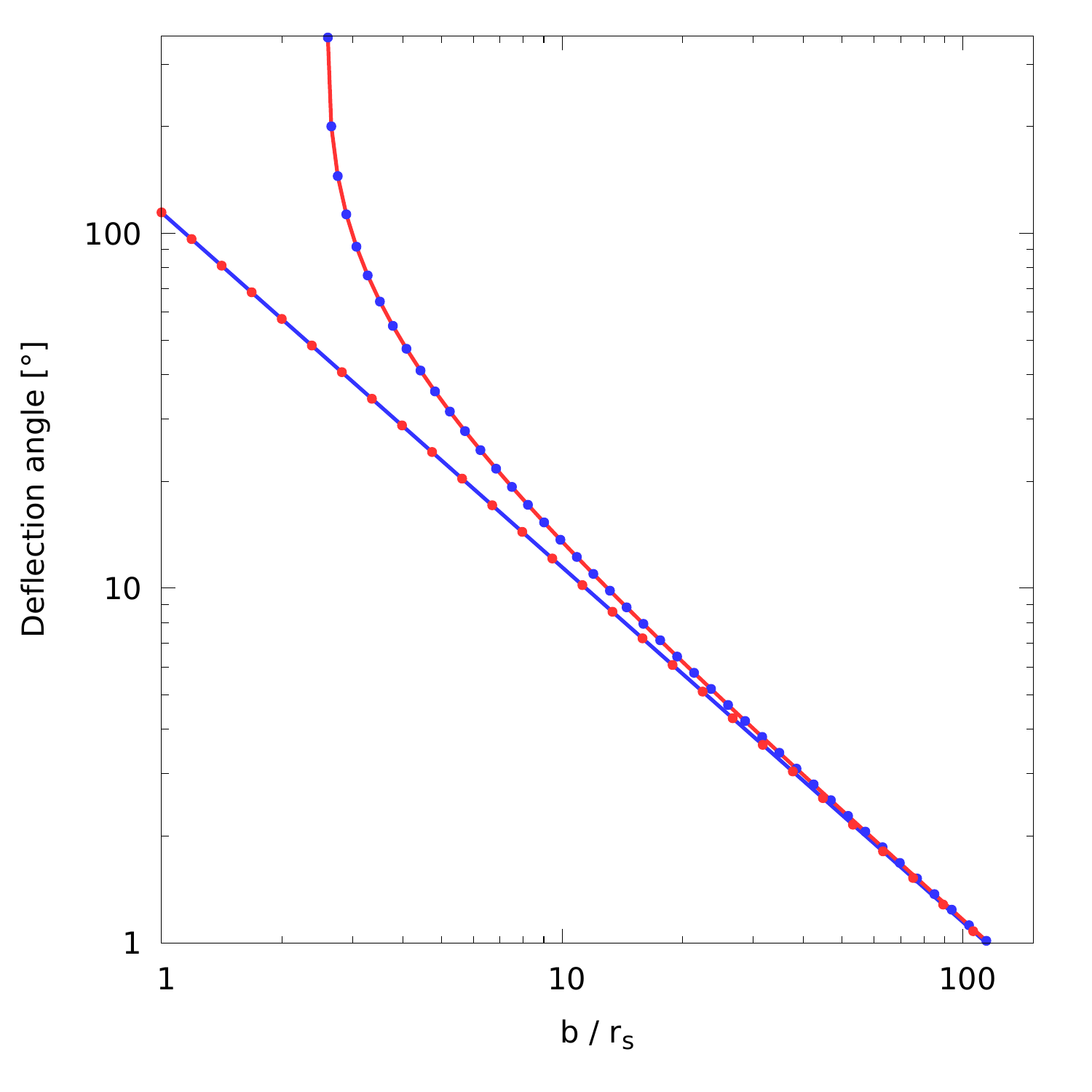}
    \includegraphics[width=0.45\textwidth]{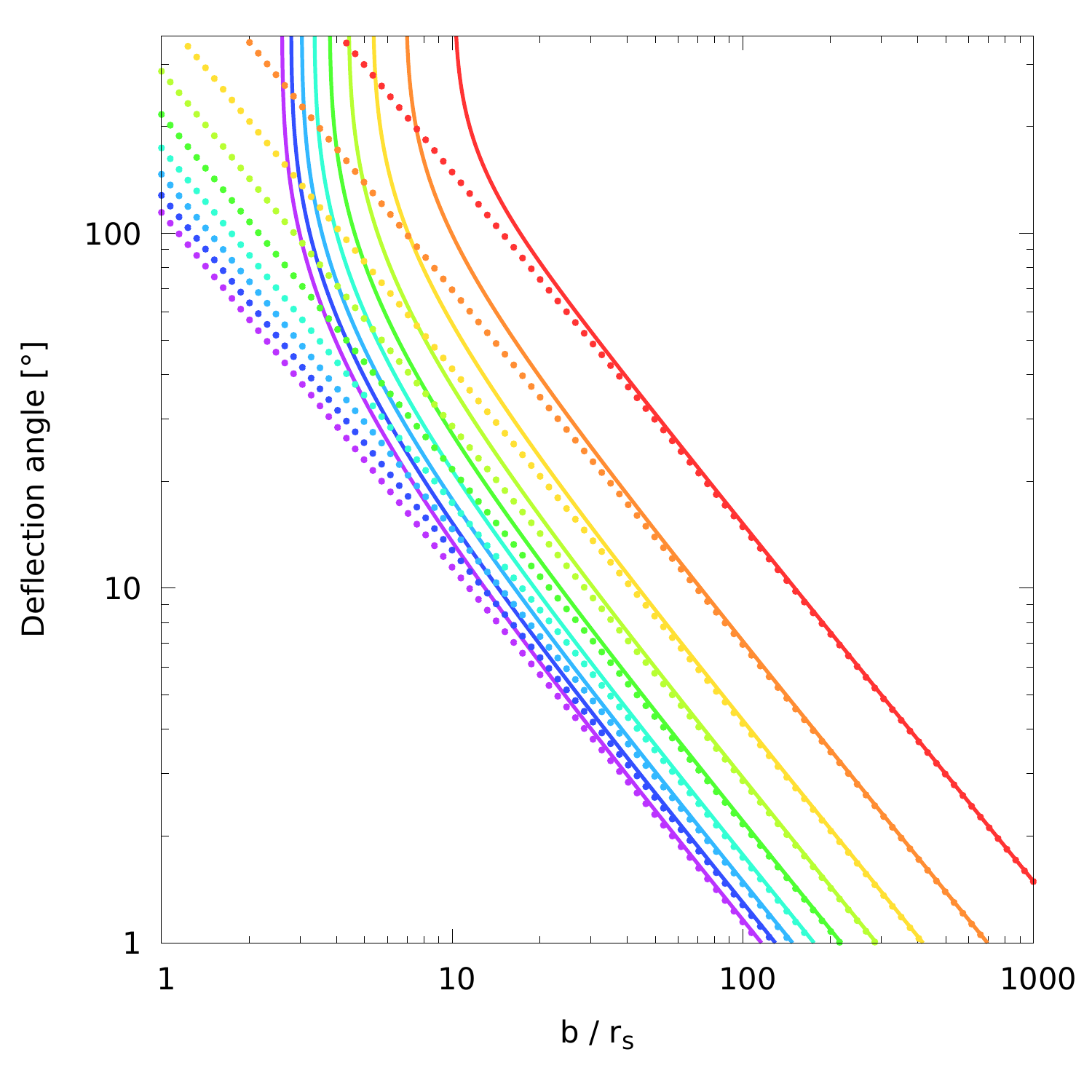}
    \caption{The left panel shows a comparison between the deflection angle for photons and  massive ultrarelativistic particles with a Lorentz factor of $100$, as a function of the normalised impact parameter $b/r_\mathrm{S}$ using the analytical limits of equations~\eqref{eq:Einstein} (massless particles, solid blue line) and \eqref{eq:analytic_deflection_masive} (massive particles, red points), the  full numerical integrations 
    of~\eqref{eq:beta_keeton} (massless particles, blue points) and \eqref{eq:deflection_masive_part} (massive particles, solid red line). 
    The right panel shows a comparison of the analytical limit of expression~\eqref{eq:analytic_deflection_masive} (dotted lines) and
    the full numerical solution~\eqref{eq:deflection_masive_part}  (solid lines) of the deflection angle of massive particles as a function of the normalised impact parameter \( b / r_\text{S} \).   From bottom to the top each colour line represents Lorentz factors of \( 100,\ 3.16,\ 2.24,\ 1.83,\ 1.58,\ 1.41,\ 1.29,\ 1.20 \) 
    and \( 1.12 \).
    }
    \label{fig:all_deflection_angles}
\end{figure*}

  Figure~\ref{fig:all_deflection_angles} shows a comparison between all these deflection 
angles for photons and massive particles as a function of the impact parameter.  The exact numerical value for the deflection angle of massive particles is the result of 
integrating numerically
equation~\eqref{eq:deflection_masive_part}, whereas for the case of light beams it is the 
result of integrating numerically equation~\eqref{eq:beta_keeton}.  

\begin{figure*}
    \centering
    \includegraphics[width=0.327\textwidth]{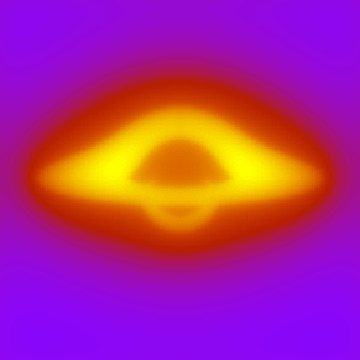}
    \includegraphics[width=0.327\textwidth]{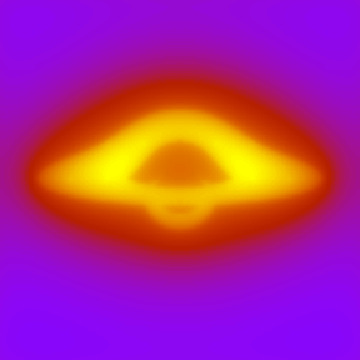}
    \includegraphics[width=0.327\textwidth]{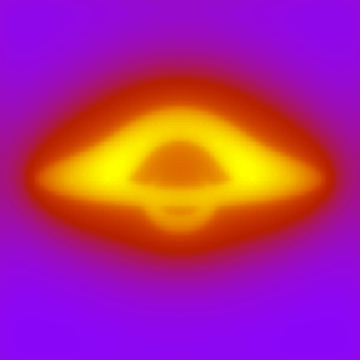}\\
    \includegraphics[width=0.327\textwidth]{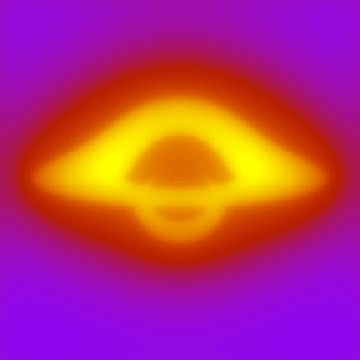}
    \includegraphics[width=0.327\textwidth]{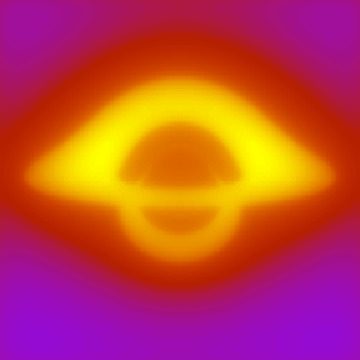}
    \includegraphics[width=0.327\textwidth]{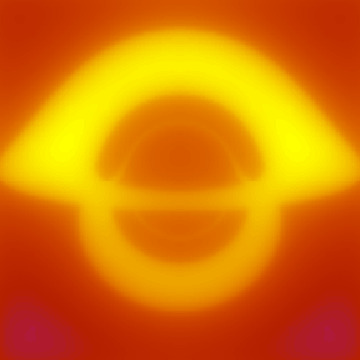}
    \caption{The Figure shows different shadows of a Schwarzschild black hole surrounded by an equatorial thin accretion disc with inner and outer radii of \( 5 \) and \( 15 \) Schwarzschild radius \( r_\text{S} \) 
    respectively. The observer is located at a 
    distance of \( 10^4 \, r_\text{S} \) and its line of sight
    makes an angle of \( 10^\circ \) with respect to the equatorial plane.    The images were generated using the free
    GNU Public Licensed software  aztekas-shadows \copyright 2024 Magallanes, Mendoza \& Santiba\~nez that 
    will be publicly available in the near future.  From left to right, top to bottom, the first panel is the shadow produced 
    by light and the remaining panels represent the shadows of 
    massive particles with Lorentz factors of \( 100,\ 3.16,\ 1.83,\ 1.41\), and \( 1.20\).}
    \label{fig:light_shadow}
\end{figure*}

\section{Discussion}

  In this article we have calculated a new set of coupled 3D spatial differential
equations~\ref{eq:general_evolution_equations} to evolve the trajectories of massless particles in any static
space-time.  We used these equations to find the deflection angle \( \alpha \) produced by a light beam on a 
Schwarzschild space-time in equation~\eqref{eq:alpha_path_integral} and compared it with the very well known 
result of the deflection angle \( \beta \) by \citet{Keeton2005Petters} shown in equation~\eqref{eq:beta_keeton}.
With this motivation, we calculated the weak deflection angle for massive particles in a Schwarzschild space-time 
and showed that for a given impact parameter the deflection angle grows as the particle's velocity decreases.  This 
fact motivates the idea as to how a black hole shadow will look if an accretion disc surrounding a central 
black hole emits massive particles (e.g. neutrinos).   Figure~\ref{fig:light_shadow} shows 
different black hole shadows of a Schwarzschild black hole surrounded by a 
thin accretion disc.   The images were produced with the use of the free GNU licensed software aztekas-shadows
\copyright 2024 Magallanes, Mendoza \& Santiba\~nez that will be publicly available in the near future.  
The software uses a 4th order Runge-Kuta integration method to solve 
the system of equations~\eqref{eq:evol_spheric_system} for the case of massless particles and the standard 
geodesic equation for the case of massive ones.  As expected, the observed image thickens with a decreasing 
particle velocity due to  greater deflection angles as compared with light.

\section{Acknowledgements}
This work was supported by a PAPIIT DGAPA-UNAM grant
IN110522. SM and MS acknowledge support from CONACyT (26344, 751147).

\bibliographystyle{aipnum4-2}
\bibliography{lensing}

\end{document}







%% file: main.bbl
\begin{thebibliography}{2}%
\makeatletter
\providecommand \@ifxundefined [1]{%
 \@ifx{#1\undefined}
}%
\providecommand \@ifnum [1]{%
 \ifnum #1\expandafter \@firstoftwo
 \else \expandafter \@secondoftwo
 \fi
}%
\providecommand \@ifx [1]{%
 \ifx #1\expandafter \@firstoftwo
 \else \expandafter \@secondoftwo
 \fi
}%
\providecommand \natexlab [1]{#1}%
\providecommand \enquote  [1]{``#1''}%
\providecommand \bibnamefont  [1]{#1}%
\providecommand \bibfnamefont [1]{#1}%
\providecommand \citenamefont [1]{#1}%
\providecommand \href@noop [0]{\@secondoftwo}%
\providecommand \href [0]{\begingroup \@sanitize@url \@href}%
\providecommand \@href[1]{\@@startlink{#1}\@@href}%
\providecommand \@@href[1]{\endgroup#1\@@endlink}%
\providecommand \@sanitize@url [0]{\catcode `\\12\catcode `\$12\catcode
  `\&12\catcode `\#12\catcode `\^12\catcode `\_12\catcode `\%12\relax}%
\providecommand \@@startlink[1]{}%
\providecommand \@@endlink[0]{}%
\providecommand \url  [0]{\begingroup\@sanitize@url \@url }%
\providecommand \@url [1]{\endgroup\@href {#1}{\urlprefix }}%
\providecommand \urlprefix  [0]{URL }%
\providecommand \Eprint [0]{\href }%
\providecommand \doibase [0]{http://dx.doi.org/}%
\providecommand \selectlanguage [0]{\@gobble}%
\providecommand \bibinfo  [0]{\@secondoftwo}%
\providecommand \bibfield  [0]{\@secondoftwo}%
\providecommand \translation [1]{[#1]}%
\providecommand \BibitemOpen [0]{}%
\providecommand \bibitemStop [0]{}%
\providecommand \bibitemNoStop [0]{.\EOS\space}%
\providecommand \EOS [0]{\spacefactor3000\relax}%
\providecommand \BibitemShut  [1]{\csname bibitem#1\endcsname}%
\let\auto@bib@innerbib\@empty
\bibitem [{\citenamefont {Schneider}, \citenamefont {Ehlers},\ and\
  \citenamefont {Falco}(2012)}]{Schneider2012gravitational}%
  \BibitemOpen
  \bibfield  {author} {\bibinfo {author} {\bibfnamefont {P.}~\bibnamefont
  {Schneider}}, \bibinfo {author} {\bibfnamefont {J.}~\bibnamefont {Ehlers}}, \
  and\ \bibinfo {author} {\bibfnamefont {E.}~\bibnamefont {Falco}},\ }\href
  {https://books.google.com.mx/books?id=WB7nBwAAQBAJ} {\emph {\bibinfo {title}
  {Gravitational Lenses}}},\ Astronomy and Astrophysics Library\ (\bibinfo
  {publisher} {Springer New York},\ \bibinfo {year} {2012})\BibitemShut
  {NoStop}%
\bibitem [{\citenamefont {Keeton}\ and\ \citenamefont
  {Petters}(2005)}]{Keeton2005Petters}%
  \BibitemOpen
  \bibfield  {author} {\bibinfo {author} {\bibfnamefont {C.~R.}\ \bibnamefont
  {Keeton}}\ and\ \bibinfo {author} {\bibfnamefont {A.~O.}\ \bibnamefont
  {Petters}},\ }\href {\doibase 10.1103/PhysRevD.72.104006} {\bibfield
  {journal} {\bibinfo  {journal} {Phys. Rev.}\ }\textbf {\bibinfo {volume}
  {D72}},\ \bibinfo {pages} {104006} (\bibinfo {year} {2005})},\ \Eprint
  {http://arxiv.org/abs/gr-qc/0511019} {arXiv:gr-qc/0511019 [gr-qc]}
  \BibitemShut {NoStop}%
\end{thebibliography}%


\begin{thebibliography}{23}%
\makeatletter
\providecommand \@ifxundefined [1]{%
 \@ifx{#1\undefined}
}%
\providecommand \@ifnum [1]{%
 \ifnum #1\expandafter \@firstoftwo
 \else \expandafter \@secondoftwo
 \fi
}%
\providecommand \@ifx [1]{%
 \ifx #1\expandafter \@firstoftwo
 \else \expandafter \@secondoftwo
 \fi
}%
\providecommand \natexlab [1]{#1}%
\providecommand \enquote  [1]{``#1''}%
\providecommand \bibnamefont  [1]{#1}%
\providecommand \bibfnamefont [1]{#1}%
\providecommand \citenamefont [1]{#1}%
\providecommand \href@noop [0]{\@secondoftwo}%
\providecommand \href [0]{\begingroup \@sanitize@url \@href}%
\providecommand \@href[1]{\@@startlink{#1}\@@href}%
\providecommand \@@href[1]{\endgroup#1\@@endlink}%
\providecommand \@sanitize@url [0]{\catcode `\\12\catcode `\$12\catcode
  `\&12\catcode `\#12\catcode `\^12\catcode `\_12\catcode `\%12\relax}%
\providecommand \@@startlink[1]{}%
\providecommand \@@endlink[0]{}%
\providecommand \url  [0]{\begingroup\@sanitize@url \@url }%
\providecommand \@url [1]{\endgroup\@href {#1}{\urlprefix }}%
\providecommand \urlprefix  [0]{URL }%
\providecommand \Eprint [0]{\href }%
\providecommand \doibase [0]{https://doi.org/}%
\providecommand \selectlanguage [0]{\@gobble}%
\providecommand \bibinfo  [0]{\@secondoftwo}%
\providecommand \bibfield  [0]{\@secondoftwo}%
\providecommand \translation [1]{[#1]}%
\providecommand \BibitemOpen [0]{}%
\providecommand \bibitemStop [0]{}%
\providecommand \bibitemNoStop [0]{.\EOS\space}%
\providecommand \EOS [0]{\spacefactor3000\relax}%
\providecommand \BibitemShut  [1]{\csname bibitem#1\endcsname}%
\let\auto@bib@innerbib\@empty
\bibitem [{\citenamefont {Dyson}, \citenamefont {Eddington},\ and\
  \citenamefont {Davidson}(1920)}]{eddington1919}%
  \BibitemOpen
  \bibfield  {author} {\bibinfo {author} {\bibfnamefont {F.~W.}\ \bibnamefont
  {Dyson}}, \bibinfo {author} {\bibfnamefont {A.~S.}\ \bibnamefont
  {Eddington}},\ and\ \bibinfo {author} {\bibfnamefont {C.}~\bibnamefont
  {Davidson}},\ }\href {http://www.jstor.org/stable/91137} {\bibfield
  {journal} {\bibinfo  {journal} {Philosophical Transactions of the Royal
  Society of London. Series A, Containing Papers of a Mathematical or Physical
  Character}\ }\textbf {\bibinfo {volume} {220}},\ \bibinfo {pages} {291}
  (\bibinfo {year} {1920})}\BibitemShut {NoStop}%
\bibitem [{\citenamefont {Mignonat}(2018)}]{Soldner1802}%
  \BibitemOpen
  \bibfield  {author} {\bibinfo {author} {\bibfnamefont {M.}~\bibnamefont
  {Mignonat}},\ }\href {https://doi.org/10.4236/jmp.2018.98095} {\bibfield
  {journal} {\bibinfo  {journal} {Journal of Modern Physics}\ }\textbf
  {\bibinfo {volume} {09}},\ \bibinfo {pages} {1545} (\bibinfo {year}
  {2018})}\BibitemShut {NoStop}%
\bibitem [{\citenamefont {Einstein}(1916)}]{einstein1916}%
  \BibitemOpen
  \bibfield  {author} {\bibinfo {author} {\bibfnamefont {A.}~\bibnamefont
  {Einstein}},\ }\href@noop {} {\  (\bibinfo {year} {1916})}\BibitemShut
  {NoStop}%
\bibitem [{\citenamefont {Luminet}(1978)}]{Luminet1978}%
  \BibitemOpen
  \bibfield  {author} {\bibinfo {author} {\bibfnamefont {J.}~\bibnamefont
  {Luminet}},\ }\href@noop {} {\bibfield  {journal} {\bibinfo  {journal}
  {Astronomy and Astrophysics}\ }\textbf {\bibinfo {volume} {75}} (\bibinfo
  {year} {1978})}\BibitemShut {NoStop}%
\bibitem [{\citenamefont {Collaboration}\ and\ \citenamefont
  {et~al.}(2022)}]{Event_Horizon_Telescope_Collaboration_2022}%
  \BibitemOpen
  \bibfield  {author} {\bibinfo {author} {\bibfnamefont {E.~H.~T.}\
  \bibnamefont {Collaboration}}\ and\ \bibinfo {author} {\bibfnamefont {K.~A.}\
  \bibnamefont {et~al.}},\ }\href {https://doi.org/10.3847/2041-8213/ac6674}
  {\bibfield  {journal} {\bibinfo  {journal} {The Astrophysical Journal
  Letters}\ }\textbf {\bibinfo {volume} {930}},\ \bibinfo {pages} {L12}
  (\bibinfo {year} {2022})}\BibitemShut {NoStop}%
\bibitem [{\citenamefont {Schneider}, \citenamefont {Ehlers},\ and\
  \citenamefont {Falco}(2012)}]{Schneider2012gravitational}%
  \BibitemOpen
  \bibfield  {author} {\bibinfo {author} {\bibfnamefont {P.}~\bibnamefont
  {Schneider}}, \bibinfo {author} {\bibfnamefont {J.}~\bibnamefont {Ehlers}},\
  and\ \bibinfo {author} {\bibfnamefont {E.}~\bibnamefont {Falco}},\ }\href
  {https://books.google.com.mx/books?id=WB7nBwAAQBAJ} {\emph {\bibinfo {title}
  {Gravitational Lenses}}},\ Astronomy and Astrophysics Library\ (\bibinfo
  {publisher} {Springer New York},\ \bibinfo {year} {2012})\BibitemShut
  {NoStop}%
\bibitem [{\citenamefont {Dodelson}(2017)}]{dodelson2017gravitational}%
  \BibitemOpen
  \bibfield  {author} {\bibinfo {author} {\bibfnamefont {S.}~\bibnamefont
  {Dodelson}},\ }\href {https://books.google.de/books?id=T3RhvgAACAAJ} {\emph
  {\bibinfo {title} {Gravitational Lensing}}}\ (\bibinfo  {publisher}
  {Cambridge University Press},\ \bibinfo {year} {2017})\BibitemShut {NoStop}%
\bibitem [{\citenamefont {Accioly}\ and\ \citenamefont
  {Ragusa}(2002)}]{A_Accioly_2002}%
  \BibitemOpen
  \bibfield  {author} {\bibinfo {author} {\bibfnamefont {A.}~\bibnamefont
  {Accioly}}\ and\ \bibinfo {author} {\bibfnamefont {S.}~\bibnamefont
  {Ragusa}},\ }\href {https://doi.org/10.1088/0264-9381/19/21/308} {\bibfield
  {journal} {\bibinfo  {journal} {Classical and Quantum Gravity}\ }\textbf
  {\bibinfo {volume} {19}},\ \bibinfo {pages} {5429} (\bibinfo {year}
  {2002})}\BibitemShut {NoStop}%
\bibitem [{\citenamefont {Pang}\ and\ \citenamefont {Jia}(2019)}]{Pang_2019}%
  \BibitemOpen
  \bibfield  {author} {\bibinfo {author} {\bibfnamefont {X.}~\bibnamefont
  {Pang}}\ and\ \bibinfo {author} {\bibfnamefont {J.}~\bibnamefont {Jia}},\
  }\href {https://doi.org/10.1088/1361-6382/ab0512} {\bibfield  {journal}
  {\bibinfo  {journal} {Classical and Quantum Gravity}\ }\textbf {\bibinfo
  {volume} {36}},\ \bibinfo {pages} {065012} (\bibinfo {year}
  {2019})}\BibitemShut {NoStop}%
\bibitem [{\citenamefont {Li}\ \emph {et~al.}(2019)\citenamefont {Li},
  \citenamefont {Zhou}, \citenamefont {Li},\ and\ \citenamefont
  {He}}]{Li_2019}%
  \BibitemOpen
  \bibfield  {author} {\bibinfo {author} {\bibfnamefont {Z.-H.}\ \bibnamefont
  {Li}}, \bibinfo {author} {\bibfnamefont {X.}~\bibnamefont {Zhou}}, \bibinfo
  {author} {\bibfnamefont {W.-J.}\ \bibnamefont {Li}},\ and\ \bibinfo {author}
  {\bibfnamefont {G.-S.}\ \bibnamefont {He}},\ }\href
  {https://doi.org/10.1088/0253-6102/71/10/1219} {\bibfield  {journal}
  {\bibinfo  {journal} {Communications in Theoretical Physics}\ }\textbf
  {\bibinfo {volume} {71}},\ \bibinfo {pages} {1219} (\bibinfo {year}
  {2019})}\BibitemShut {NoStop}%
\bibitem [{\citenamefont
  {Deppisch}(2019)}]{A_Modern_Introduction_to_Neutrino_Physics}%
  \BibitemOpen
  \bibfield  {author} {\bibinfo {author} {\bibfnamefont {F.~F.}\ \bibnamefont
  {Deppisch}},\ }\href {https://doi.org/10.1088/2053-2571/ab21c9} {\emph
  {\bibinfo {title} {A Modern Introduction to Neutrino Physics}}},\ 2053-2571\
  (\bibinfo  {publisher} {Morgan \& Claypool Publishers},\ \bibinfo {year}
  {2019})\BibitemShut {NoStop}%
\bibitem [{Ice()}]{Icecube}%
  \BibitemOpen
  \href {https://masterclass.icecube.wisc.edu/es/aprende/deteccion_neutrinos}
  {\emph {\bibinfo {title} {Icecube}}}\BibitemShut {NoStop}%
\bibitem [{\citenamefont {Surman}\ and\ \citenamefont
  {Mclaughlin}(2003)}]{Rebecca_Surman_et.al.}%
  \BibitemOpen
  \bibfield  {author} {\bibinfo {author} {\bibfnamefont {R.}~\bibnamefont
  {Surman}}\ and\ \bibinfo {author} {\bibfnamefont {G.}~\bibnamefont
  {Mclaughlin}},\ }\href {https://doi.org/10.1086/381672} {\bibfield  {journal}
  {\bibinfo  {journal} {The Astrophysical Journal}\ }\textbf {\bibinfo {volume}
  {603}} (\bibinfo {year} {2003}),\ 10.1086/381672}\BibitemShut {NoStop}%
\bibitem [{\citenamefont {Nellen}, \citenamefont {Mannheim},\ and\
  \citenamefont {Biermann}(1993)}]{PhysRevD.47.5270}%
  \BibitemOpen
  \bibfield  {author} {\bibinfo {author} {\bibfnamefont {L.}~\bibnamefont
  {Nellen}}, \bibinfo {author} {\bibfnamefont {K.}~\bibnamefont {Mannheim}},\
  and\ \bibinfo {author} {\bibfnamefont {P.~L.}\ \bibnamefont {Biermann}},\
  }\href {https://doi.org/10.1103/PhysRevD.47.5270} {\bibfield  {journal}
  {\bibinfo  {journal} {Phys. Rev. D}\ }\textbf {\bibinfo {volume} {47}},\
  \bibinfo {pages} {5270} (\bibinfo {year} {1993})}\BibitemShut {NoStop}%
\bibitem [{\citenamefont {Landau}\ and\ \citenamefont
  {Lifshitz}(2013)}]{landau2013classical}%
  \BibitemOpen
  \bibfield  {author} {\bibinfo {author} {\bibfnamefont {L.}~\bibnamefont
  {Landau}}\ and\ \bibinfo {author} {\bibfnamefont {E.}~\bibnamefont
  {Lifshitz}},\ }\href {https://books.google.com.mx/books?id=HudbAwAAQBAJ}
  {\emph {\bibinfo {title} {The Classical Theory of Fields}}},\ COURSE OF
  THEORETICAL PHYSICS\ (\bibinfo  {publisher} {Elsevier Science},\ \bibinfo
  {year} {2013})\BibitemShut {NoStop}%
\bibitem [{\citenamefont {Misner}\ \emph {et~al.}(2017)\citenamefont {Misner},
  \citenamefont {Thorne}, \citenamefont {Wheeler},\ and\ \citenamefont
  {Kaiser}}]{misner2017gravitation}%
  \BibitemOpen
  \bibfield  {author} {\bibinfo {author} {\bibfnamefont {C.}~\bibnamefont
  {Misner}}, \bibinfo {author} {\bibfnamefont {K.}~\bibnamefont {Thorne}},
  \bibinfo {author} {\bibfnamefont {J.}~\bibnamefont {Wheeler}},\ and\ \bibinfo
  {author} {\bibfnamefont {D.}~\bibnamefont {Kaiser}},\ }\href
  {https://books.google.com.mx/books?id=SyQzDwAAQBAJ} {\emph {\bibinfo {title}
  {Gravitation}}}\ (\bibinfo  {publisher} {Princeton University Press},\
  \bibinfo {year} {2017})\BibitemShut {NoStop}%
\bibitem [{\citenamefont {Poisson}\ and\ \citenamefont
  {Will}(2014)}]{Poisson2014gravity}%
  \BibitemOpen
  \bibfield  {author} {\bibinfo {author} {\bibfnamefont {E.}~\bibnamefont
  {Poisson}}\ and\ \bibinfo {author} {\bibfnamefont {C.}~\bibnamefont {Will}},\
  }\href {https://books.google.com.mx/books?id=PZ5cAwAAQBAJ} {\emph {\bibinfo
  {title} {Gravity: Newtonian, Post-Newtonian, Relativistic}}}\ (\bibinfo
  {publisher} {Cambridge University Press},\ \bibinfo {year}
  {2014})\BibitemShut {NoStop}%
\bibitem [{\citenamefont {Hobson}\ \emph {et~al.}(2006)\citenamefont {Hobson},
  \citenamefont {P}, \citenamefont {Efstathiou}, \citenamefont {Lasenby},\ and\
  \citenamefont {Press}}]{hobson2006general}%
  \BibitemOpen
  \bibfield  {author} {\bibinfo {author} {\bibfnamefont {M.}~\bibnamefont
  {Hobson}}, \bibinfo {author} {\bibfnamefont {E.}~\bibnamefont {P}}, \bibinfo
  {author} {\bibfnamefont {G.}~\bibnamefont {Efstathiou}}, \bibinfo {author}
  {\bibfnamefont {A.}~\bibnamefont {Lasenby}},\ and\ \bibinfo {author}
  {\bibfnamefont {C.~U.}\ \bibnamefont {Press}},\ }\href
  {https://books.google.com.mx/books?id=5dryXCWR7EIC} {\emph {\bibinfo {title}
  {General Relativity: An Introduction for Physicists}}},\ General Relativity:
  An Introduction for Physicists\ (\bibinfo  {publisher} {Cambridge University
  Press},\ \bibinfo {year} {2006})\BibitemShut {NoStop}%
\bibitem [{\citenamefont {{Younsi}}(2014{\natexlab{a}})}]{Ziri_Younsi}%
  \BibitemOpen
  \bibfield  {author} {\bibinfo {author} {\bibfnamefont {Z.}~\bibnamefont
  {{Younsi}}},\ }\emph {\bibinfo {title} {{General relativistic radiative
  transfer in black hole systems}}},\ \href@noop {} {Ph.D. thesis},\ \bibinfo
  {school} {University College London, UK} (\bibinfo {year}
  {2014}{\natexlab{a}})\BibitemShut {NoStop}%
\bibitem [{\citenamefont {{Younsi}}(2014{\natexlab{b}})}]{2014PhDT_Younsi}%
  \BibitemOpen
  \bibfield  {author} {\bibinfo {author} {\bibfnamefont {Z.}~\bibnamefont
  {{Younsi}}},\ }\emph {\bibinfo {title} {{General relativistic radiative
  transfer in black hole systems}}},\ \href@noop {} {Ph.D. thesis},\ \bibinfo
  {school} {University College London, UK} (\bibinfo {year}
  {2014}{\natexlab{b}})\BibitemShut {NoStop}%
\bibitem [{\citenamefont {Keeton}\ and\ \citenamefont
  {Petters}(2005)}]{Keeton2005Petters}%
  \BibitemOpen
  \bibfield  {author} {\bibinfo {author} {\bibfnamefont {C.~R.}\ \bibnamefont
  {Keeton}}\ and\ \bibinfo {author} {\bibfnamefont {A.~O.}\ \bibnamefont
  {Petters}},\ }\href {https://doi.org/10.1103/PhysRevD.72.104006} {\bibfield
  {journal} {\bibinfo  {journal} {Phys. Rev.}\ }\textbf {\bibinfo {volume}
  {D72}},\ \bibinfo {pages} {104006} (\bibinfo {year} {2005})},\ \Eprint
  {https://arxiv.org/abs/gr-qc/0511019} {arXiv:gr-qc/0511019 [gr-qc]}
  \BibitemShut {NoStop}%
\bibitem [{GNU(2019)}]{GNU_Scientific_Libraries}%
  \BibitemOpen
  \href {https://www.gnu.org/software/gsl/doc/html} {\emph {\bibinfo {title}
  {GSL}}} (\bibinfo {year} {1996-2019})\BibitemShut {NoStop}%
\bibitem [{\citenamefont {Tejeda}\ and\ \citenamefont
  {Aguayo-Ortiz}(2019)}]{Tejeda_and_Aguayo2019}%
  \BibitemOpen
  \bibfield  {author} {\bibinfo {author} {\bibfnamefont {E.}~\bibnamefont
  {Tejeda}}\ and\ \bibinfo {author} {\bibfnamefont {A.}~\bibnamefont
  {Aguayo-Ortiz}},\ }\href {https://doi.org/10.1093/mnras/stz1513} {\bibfield
  {journal} {\bibinfo  {journal} {Monthly Notices of the Royal Astronomical
  Society}\ }\textbf {\bibinfo {volume} {487}},\ \bibinfo {pages} {3607}
  (\bibinfo {year} {2019})},\ \Eprint
  {https://arxiv.org/abs/https://academic.oup.com/mnras/article-pdf/487/3/3607/28844494/stz1513.pdf}
  {https://academic.oup.com/mnras/article-pdf/487/3/3607/28844494/stz1513.pdf}
  \BibitemShut {NoStop}%
\end{thebibliography}%
